\documentclass{article}

\usepackage{arxiv}

\usepackage[utf8]{inputenc} 
\usepackage[T1]{fontenc}    
\usepackage{hyperref}       
\usepackage{url}            
\usepackage{booktabs}       
\usepackage{amsfonts}       
\usepackage{nicefrac}       
\usepackage{microtype}      
\usepackage{lipsum}
\usepackage{graphicx}

\usepackage{amsmath, amsfonts, amssymb, amsthm}
\usepackage{subcaption}
\usepackage{empheq}
\usepackage{multirow}
\usepackage{cleveref}

\newcommand*{\E}{\mathbb{E}}
\newcommand*{\Prb}{\mathbb{P}}
\newcommand*{\dottheta}{\Dot{\theta}}
\newcommand*{\dotpop}{\Dot{\alpha}_1}
\newcommand*{\dotniche}{\Dot{\alpha}_{-1}}
\newcommand*{\apop}{\alpha_1}
\newcommand*{\aniche}{\alpha_{-1}}
\newcommand*{\ppop}{\hat{p}_1}
\newcommand*{\pniche}{\hat{p}_{-1}}

\newcommand*{\mupop}{\mu_1}
\newcommand*{\muniche}{\mu_{-1}}
\newcommand*{\spop}{\Sigma_1}
\newcommand*{\sniche}{\Sigma_{-1}}
\newcommand*{\tstar}{\theta^\ast}

\newtheorem{theorem}{Theorem}[section]
\newtheorem{lemma}[theorem]{Lemma}

\newtheorem{assumption}{Assumption}

\title{Stay or Stray -- A Dynamical Systems Viewpoint of Popularity Bias}

\author{
Sarvesh Shashidhar \\
Centre of Machine Intelligence \& Data Science \\
Indian Institute of Technology, Bombay \\
Mumbai, Maharashtra, India \\
\texttt{ssarvesh@iitb.ac.in} \\
\And
Lankireddy Prabhat \\
Electrical \& Computer Engineering Dept. \\
Georgia Institute of Technology \\
Atlanta, GA, USA \\
\texttt{plankireddy3@gatech.edu} \\
\And
Arpit Agarwal \\
Dept. of Computer Science \& Engineering\\
Indian Institute of Technology, Bombay \\
Mumbai, Maharashtra, India \\
\texttt{aarpit@iitb.ac.in} \\
\And
D.  Manjunath \\
Dept. of Electrical Engineering\\
Indian Institute of Technology, Bombay \\
Mumbai, Maharashtra, India \\
\texttt{dmanju@iitb.ac.in} \\
\And
Karan Bhukar \\
Amazon Music \\
Bengaluru, Karnataka, India \\
\texttt{bhukar@amazon.com} \\
\And
Tanmay Khandelwal \\
Amazon Music \\
Sunnyvale, California, USA \\
\texttt{tanmayx@amazon.com} \\
}

\begin{document}
\maketitle
\begin{abstract}
Popularity bias in recommendation systems arises when a \emph{majority} user class generates disproportionate interaction data, causing the system to increasingly favour it while degrading recommendation quality for \emph{niche} users. While extensive empirical evidence of popularity bias exists, the \emph{dynamics} leading to its emergence are not well understood.
In this work, we study the \emph{coupled evolution} of recommender model updates and user engagement through the lens of dynamical systems. We formulate a stochastic process and analyse its asymptotic behaviour through an ordinary differential equation (ODE) framework grounded in two-time-scale stochastic approximation. We characterise the equilibrium points of this dynamical system, and derive conditions under which popularity bias is provably emergent, as well as conditions under which symmetric retention of all user classes is possible. We conduct experiments on synthetic data and real-world production logs derived from a large-scale commercial music recommendation platform to validate our theoretical results.
\end{abstract}


\section{Introduction}
Modern recommendation systems (RS) have become integral to digital media platforms \cite{realworld2024survey}, mediating the interaction between users and items by learning user preferences from observed behaviour. The effectiveness of these systems relies on the accurate modelling of user preferences, typically achieved through online learning on interaction data \cite{survey2023popularity, coldstart2025inherited}. In practice, however, user populations are rarely balanced \cite{park2008longtail, celma2010longtail}-- a dominant user class contributes a disproportionate share of interactions, leading the system to preferentially learn majority preferences at the expense of niche groups. This phenomenon, commonly known as \emph{popularity bias} \cite{bhadani2021biases}, has been well documented in the literature \cite{diversity2017novelty, beel2013impact, abdollahpouri2019managing, he2022mitigating}. 

However, the vast majority of existing work examines popularity bias through a static lens, capturing only a snapshot of the bias and studying its consequences without taking into account the underlying mechanisms that generate and amplify it over time \cite{diversity2017novelty, beel2013impact}. In this work, we take a dynamical systems perspective to study the emergence of popularity bias in recommendation systems and its downstream effects on niche user churn.

We formulate a model where there are two classes of users: a majority (\emph{popular}) class and a minority (\emph{niche}) class. The recommender system is an online learner that updates its parameters to better match the user class with its preferred content, while users update their rates of arrival based on preference match.\footnote{The change in user arrival rates in response to RS quality is motivated by our empirical study from a commercial music RS, where we found high correlation between user churn rates and their preferences being matched (see \Cref{sec:expts}).}  To understand the long-term dynamics of this coupled system, we employ a continuous-time ordinary differential equation (ODE) framework grounded in two-timescale stochastic approximation \cite{borkar2008stochastic}. Specifically, we model the system as evolving on two different timescales: the recommendation algorithm adapts rapidly with each user interaction, while the user population is slow in adjusting its arrival rate.\footnote{This is because user preferences tend to exhibit tolerance towards sub-optimal recommendations, and shift only when there is sustained drop in recommendation quality \cite{COPPOLILLO2025104125}.}
This two-timescale analysis permits treating the arrival rates as quasi-static relative to model parameter changes. 
Substituting the optimal model parameters for the current arrival rates yields a simplified ODE over arrival rates alone, enabling closed-form analysis of long-term equilibrium behavior.

We characterise the equilibrium structure and convergence properties of this reduced system. Under a non-collinearity condition on the class-conditional feature means, we show that the four corners of $[0,1]^2$ are valid equilibrium points, each admitting a natural interpretation-- $(1,1)$ corresponds to retention of both user classes, $(1,0)$ and $(0,1)$ correspond to selective disengagement of one class (i.e., popularity bias), and $(0,0)$ corresponds to simultaneous disengagement. Our analysis yields three principal theoretical results. First, we prove that the equilibrium $(0,0)$ is not asymptotically reachable from any interior initial condition, thereby ruling out simultaneous disengagement as a feasible long-run outcome (\Cref{thm:4_1}). Second, we derive a threshold $p^\ast$ on the fraction of majority users such that when $p > p^\ast$, the system asymptotically converges to $(1,0)$---the niche class disengages entirely while the majority class is retained---formally characterising the onset of popularity bias (\Cref{thm:4_2}). Third, we identify two complementary regimes that guarantee \emph{symmetric retention}, i.e., convergence to $(1,1)$-- a sufficient condition based on the sign of the inner product of the class-conditional means in a covariance-transformed space (\Cref{thm:4_3}), and an existence result that identifies a feasible interval of $p$ values ensuring both classes are retained, even when the sufficient condition of \Cref{thm:4_3} does not hold (\Cref{thm:4_4}).

We validate our theoretical predictions through simulations on synthetic data, demonstrating that the system trajectories conform to the equilibrium behaviour predicted by our theorems across a broad range of initialisations and parameter configurations. Additionally, we construct a semi-synthetic evaluation using production engagement logs from a large-scale commercial music recommendation platform comprising approximately $410$ million user--item interactions. By fitting our model to user embeddings derived from matrix factorisation on this interaction data, we show that the observed platform dynamics---including disproportionate churn among niche-preferring users---are consistent with the conditions identified by our framework.
Our theoretical framework coupled with empirical evidence suggests a natural mitigation strategy that focuses on balancing the accuracy of the user classes which is similar in spirit to \cite{yu2024fairbalance}. Our work provides additional justification for this strategy and we empirically demonstrate its effectiveness in reducing popularity bias. 

\section{Related Work}\label{sec:related}
Our work is closely related to the literature on popularity bias and dynamical system analysis. Extensive prior research has characterised popularity bias \cite{park2008longtail} and the degradation of novelty, diversity, and user satisfaction in the long-tail  \cite{celma2010longtail, beel2013impact, bhadani2021biases}. This has resulted in multiple interventions being proposed at the pre-processing \cite{gulsoy2025equirate}, in-processing \cite{itemloss2024correcting,zhang2021causal} and post-processing \cite{abdollahpouri2019managing} stages. These strategies predominantly address bias symptoms rather than the underlying mechanisms. Concurrently, dynamic RS frameworks have explored user preference drift and exposure effects \cite{COPPOLILLO2025104125, rafailidis2015modeling}. However, they typically treat user engagement as an exogenous variable, and the co-evolution of user behaviour and model parameters remains largely unexplored. A detailed section on related works has been provided in supplementary Section A.

\section{Dynamic Model Formulation} \label{sec:model}
 
\paragraph{Binary user-classes.}
We consider a model consisting of two user groups-- class $+1$ (majority/popular users) and class $-1$ (minority/niche users). The items are similarly partitioned into two classes indexed by $\{-1, +1\}$.
\begin{assumption}
    \label{ass:1}
    The users of class $c_u \in \{-1,+1\}$ prefer the items belonging to class $c_i \in \{-1,+1\}$ if $c_u = c_i$. As a result, the user utility reduced to the bilinear form $$ \text{Utility}(u,i) = c_u \cdot c_i$$
    where $c_u$ and $c_i$ represent the classes of the user and item respectively.
\end{assumption}

Under \cref{ass:1}, user preferences are perfectly aligned with class identity. Consequently, the task of providing recommendations reduces to learning a binary classifier over user features. 
This simplification makes our analysis mathematically tractable; the extension of the results to a multi-class system follows naturally and has been detailed in Section B in the supplementary material.

\paragraph{Arrival dynamics.}
We now formalize the notation governing the dynamics. Let \(p \in (0,1)\) denote the fraction of users belonging to class $+1$. User arrival into the system is stochastic -- at time $t$, a class $+1$ user enters the system with probability \((\apop)_t \in [0,1]\), while a niche user enters with probability \((\aniche)_t \in [0,1]\). Quantities $(\apop)_t$ and $(\aniche)_t$, also called \emph{walk-up rates}, combined with user class imbalance gives us the overall probability of the arrival of class $+1$ users and class $-1$ users as $p (\apop)_t$ and $(1 - p)(\aniche)_t$ respectively. With the remaining probability, no user arrives to the system (equivalent to having $y_t = 0$). For notational convenience, we shall define $\alpha_t = ((\apop)_t, (\aniche)_t)$ as a 2-tuple of walk-up rates. 

\paragraph{User features.}
The user arriving at time $t$ is represented using a feature vector \(X_t \in \mathbb{R}^d\) and class label \(y_t \in \{-1,+1\}\). The users are then classified using an online binary classifier, parameterised by \(\theta_t \in \mathbb{R}^d\) which is learnable. The classifier parameter $(\theta_t)$ and the walk-up rates of the users $((\apop)_t, (\aniche)_t)$ {evolve jointly} over time, with the resulting coupled dynamics dictating the long-term user behaviour. The user preferences (which remain static with time) aid in characterizing the extent of class imbalance, i.e., value of $p$.

\begin{assumption}
    \label{ass:2}
    The feature vectors of the users, $X_t \in \mathbb{R}^{d},$ are sampled from class-conditional Gaussian distributions. 
    \[
        X_t |_{y_t = 1} \sim \mathcal{N}(\mupop, \spop) \quad ; \quad X_t |_{y_t = -1} \sim \mathcal{N}(\muniche, \sniche)
    \]
\end{assumption}

Imposing Gaussian priors on user representations is a well-known technique \cite{ruslan2008probabilistic, liang2018vaecf} that allows for tractable analysis. In our case, it allows for a closed-form expression for the optimal classifier threshold $\tstar$ going forward. In our experimental analysis in \Cref{sec:rw} and supplementary Section J.5, we show that the theoretical trends hold for real-world data even though it is not perfectly Gaussian in nature.

\paragraph{Parameter updates.} 
We now outline the discrete time operation of the RS, which is an online binary classifier. At each time step $t$, a user with feature vector $X_t$ is classified by the model using the rule $\hat{y}_t = \text{sign}\left( \theta_t^\top X_t \right)$. If $\hat{y}_t = y_t$ (correctly classified), then the users get serviced optimally by the RS and they experience an increase in their walk-up rates $((\alpha_{y_t})_{t+1} > (\alpha_{y_t})_t)$, leading to an overall increase in their arrival. On the other hand, a misclassification ($\hat{y}_t \neq y_t$) results in a decrease in walk-up rates $((\alpha_{y_t})_{t+1} < (\alpha_{y_t})_t)$. After each update, the model updates its parameters, i.e. $\theta_t$ to ensure lower misclassification rates.

We now derive the continuous-time approximation of the discrete stochastic update mentioned above using \emph{ordinary differential equations (ODEs)}. Let us first characterize the evolution of $\theta_t$, which is updated by the RS to minimize the \emph{regularised mean-squared error (MSE)} defined as follows --

\begin{align*}
    \ell (\theta_t; X_t, y_t) & = \frac{1}{2} [y_t - \theta_t^\top X_t]^2 + \frac{\lambda}{2} \Vert \theta_t \Vert^2 \\
    L (\theta_t, \alpha_t) & = \E \left[ \ell(\theta_t; X_t, y_t) | \theta_t, \alpha_t \right]
\end{align*}

While it is natural to use either hinge or log loss for online binary classifiers, recent work \cite{arora2019mse} has shown that MSE loss also empirically performs well. Additionally, using regularised MSE loss allows for a closed-form expression for $\tstar$. However, we perform experiments with other loss functions in supplementary Section C to show that our theoretical insights are valid beyong the MSE loss.

The value of $\theta_{t}$ is updated using \emph{online gradient descent} over $L(\theta_t)$, which can be viewed as a discretisation of the \emph{gradient flow} ODE \cite{borkar2008stochastic}.
\begin{equation}
    \label{eq:dot_t}
    \begin{split}
        \theta_{t+1} & = \theta_t - \eta_t \nabla_\theta \left( \ell(\theta_t) \right) \quad \equiv \quad \dottheta = - \nabla_\theta \left( L(\theta, \alpha) \right)
    \end{split}
\end{equation}

Now we characterize the update for the walk-up rates, starting with $(\apop)_t$. The value of $(\apop)_t$ either increases ($\hat{y}_t = y_t = 1$), decreases ($\hat{y}_t \neq y_t = 1$) or doesn't change ($y_t = -1$). These three conditions can be combined to give a single expression for updating $(\apop)_t$.
\begin{equation*}
    (\apop)_{t+1} = (\apop)_{t} + \delta_t \left[ \text{sign}(\theta_t^\top X_t) \cdot \left( \frac{1 + y_t}{2} \right) \right]
\end{equation*}

where $\delta_t$ is the time-decaying step-size to ensure asymptotic convergence. Taking conditional expectation over $X_t$ yields the continuous time stochastic update for $(\apop)_t$.
{
\begin{equation*}
    \begin{split}
        & \frac{1}{\delta_t} \E \left[ (\apop)_{t+1} - (\apop)_{t} \right] = \E\left[ \text{sign}(\theta_t^\top X_t) \cdot \left( \frac{1 + y_t}{2} \right) \right] \\
        & = \Prb[y_t = 1] \left\{ \Prb_{X_t \vert y_t=1} [\theta_t^\top X_t \geq 0] - \Prb_{X_t \vert y_t=1} [\theta_t^\top X_t < 0] \right\} \\
        & = p (\apop)_t \left\{ 2 (\ppop)_t - 1 \right\}
    \end{split}
\end{equation*}
}
where $(\ppop)_t = \Prb_{X_t \vert y_t=1} [\theta_t^\top X_t \geq 0]$ is the \emph{True Positive Rate (TPR)} of the classifier at time $t$. Applying \emph{projected} stochastic approximation yields the ODE for $(\apop)_t$. 
\begin{equation}
    \label{eq:dot_1}
    \dotpop = \Gamma \left\{ p \apop [2\ppop - 1] \right\}
\end{equation}

where $\Gamma (x)$ is a projection operator that restricts $x$ to the compact set $[0, 1]$. Similarly, the update for $(\aniche)_t$ can also be approximated to a continuous-time ODE. 
\begin{align}
    \label{eq:dot_2}
    (\aniche)_{t+1} & = (\aniche)_t + \delta_t \left[ - \text{sign}(\theta_t^\top X_t) \cdot \left( \frac{1 - y_t}{2} \right) \right] \nonumber \\
    \dotniche & = \Gamma \left\{ (1 - p) \aniche [2\pniche - 1] \right\}
\end{align}

Where $(\pniche)_t = \Prb_{X_t \vert y_t=-1} [\theta_t^\top X_t < 0]$ is the \emph{True Negative Rate (TNR)} of the classifier at time $t$. Equations \ref{eq:dot_t}, \ref{eq:dot_1} \& \ref{eq:dot_2} characterize the coupled dynamics of model parameters ($\theta_t$) and user behaviour ($\alpha_t$).

\subsection{Two-Time-Scale Stochastic Approximation} \label{sec:tts}
In a practical RS, the users (at an aggregate level) have inertia and tolerance \cite{COPPOLILLO2025104125} towards sub-optimal recommendations, while the parameters of the RS are updated    much more frequently. Essentially, the user arrival rates are expected to evolve \emph{at a slower time scale} when compared to the model parameters.

\begin{assumption}
    \label{ass:3}
    Consider the stochastic updates --
    \begin{align*}
        \theta_{t+1} & = \theta_t - \eta_t \nabla_\theta \ell(\theta_t; X_t, y_t) \\
        (\apop)_{t+1} & = (\apop)_t + \delta_t H_1 (\theta_t, X_t, y_t) \\
        (\aniche)_{t+1} & = (\aniche)_t + \delta_t H_{-1} (\theta_t, X_t, y_t)
    \end{align*}
    where the mean-drifts for each quantity---$\nabla_\theta L, \E[H_1], \E[H_2]$---are Lipschitz. Additionally, the decaying step-sizes---$\eta_t, \delta_t$ -- are such that $\sum_{t=0}^{\infty} \eta_t = \infty$, $\sum_{t=0}^{\infty} \delta_t = \infty$, while both $\sum_{t=0}^{\infty} \eta_t^2$ and $\sum_{t=0}^{\infty} \delta_t^2$ are finite. Then, we assume that $ \lim_{t \to \infty} \, \frac{\delta_t}{\eta_t} = 0$ i.e. $\theta_t$ evolves faster than $\alpha_t$.
\end{assumption}

Not only does \Cref{ass:3} highlight the relation between user and model updates observed in practice, it also is the foundational idea on which we build our framework. With the appropriate choice of $\delta_t$ and $\eta_t$ (as provided by \cite{borkar2008stochastic}), we can effectively treat the faster transient ($\alpha_t$) as \emph{quasi-static} with respect to the slower transient ($\theta_t$). This allows us to find an expression for $\tstar$ in terms of $\alpha_t$. Note that if we flip the time scales, we might end up in a situation where the users update their behaviour faster than the system can learn their preferences, which is not ideal. With $\alpha_t = \alpha$ (static), we write the expression for $\tstar$ as follows.

\begin{equation}
    \label{eq:tstar}
    \begin{split}
        & \tstar(\alpha) = F(\alpha) = A_\alpha^{-1} b_\alpha \\
        A_\alpha & = p \apop (\spop + \mupop \mupop^\top) \\
        & + (1 - p)\aniche (\sniche + \muniche \muniche^\top ) + \lambda I \\
        b_\alpha & = p \apop \mupop - (1 - p) \aniche \muniche
    \end{split}
\end{equation}

This expression for $\tstar$ is the minimizer of $L(\theta_t)$ and is found by setting $\nabla_\theta L(\theta_t) = 0$. The derivation of \Cref{eq:tstar} along with a proof for Lipschitzness of $F$ (using first principles) has been deferred to Section E in the supplementary material. Crucially, thanks to \Cref{ass:3}, our framework has reduced to a two-ODE system of just the walk-up rates. This elegant reduction using two-time scale separation has been explored by \cite{borkar2008stochastic}, the details of which have been highlighted in Section D in the supplementary material. After treating $\alpha_t$ as static and getting $\tstar$, the framework then uses $\tstar$ to update $\alpha_t$. The cycle continues and this gives rise to their coupled evolution.

\subsection{Characterizing System Equilibria} \label{sec:eq_points}
Using the two-ODE framework defined so far, we aim to find the limiting behaviour of the users and the system, which essentially boils down to finding the equilibrium points where the system converges to asymptotically. Any point $\alpha^\ast = (\apop^\ast, \aniche^\ast) \in [0,1]^2$ is said to be an equilibrium point of the system if $\dotpop\vert_{\alpha^\ast} = \dotniche\vert_{\alpha^\ast} = 0$, which also results in $\dottheta\vert_{\alpha^\ast} = 0$.

Prior to identifying the equilibrium points, we use \Cref{ass:2} to characterise the evolution of the walk-up rates in terms of means of the user feature distributions.

\begin{lemma}
    \label{lem:1}
    For all $\alpha^\ast \in [0,1]^{2}$ and $\tstar = F(\alpha^\ast)$, we see that $\dotpop\vert_{\alpha^\ast} > 0$ and $\dotniche\vert_{\alpha^\ast} > 0$ are equivalent to having $(\tstar)^\top \mupop > 0$ and $(\tstar)^\top \muniche < 0$ respectively.
\end{lemma}

\paragraph{Proof sketch.}

The lemma above follows from the fact that for user class $i$, the true rate $\hat{p}_i$ depends (by definition) on the value of $\theta^\top X$ which is a random variable of the Gaussian distribution $\mathcal{N}(\theta^\top  \mu_i, \theta^\top \Sigma_i \theta)$. Performing simple substitution, as highlighted in detail in supplementary Section F, proves Lemma~\ref{lem:1}. It must be noted that Lemma~\ref{lem:1} represents the \emph{necessary \& sufficient} conditions for an increase in walk-up rates.

From Lemma~\ref{lem:1}, we see that $\dotpop = 0$ is equivalent to $\theta^\top \mupop = 0$. Now suppose if the means of the distributions are collinear i.e. $\exists c \in \mathbb{R}$ such that $\muniche = c \cdot \mupop$, then $\theta^\top \mupop = 0$ is equivalent to $\theta^\top \muniche = 0$ which implies $\dotniche = 0$. Hence, equilibrium analysis for collinear means becomes trivial, and we choose not to focus on this degenerate case.

\begin{assumption}
    \label{ass:4}
    Consider the means--$\mupop, \muniche$ of the distribution of user features. Then $\nexists \hspace{0.1cm} c \in \mathbb{R}$ such that $\mupop = c \cdot \muniche$.
\end{assumption}

Now we can proceed to find the equilibrium points.

\begin{lemma}
    \label{lem:2}
    Under Assumptions~\ref{ass:1}--\ref{ass:4}, the set of equilibria is given by $\mathcal{E} = \left\{ (0,0), (1,x), (y, 1) \right\}$ where $x = \frac{p \langle \muniche, \mupop \rangle_{A^{-1}}}{(1 - p) \Vert \muniche \Vert_{A^{-1}}^{2}}$ and $y = \frac{(1 - p) \langle \muniche, \mupop \rangle_{A^{-1}}}{p \Vert \mupop \Vert_{A^{-1}}^{2}}$.
\end{lemma}

\paragraph{Proof sketch}
The first step is to set Equations \ref{eq:dot_1} and \ref{eq:dot_2} to $0$. Next, by checking the conditions of Equations \ref{eq:dot_1} and \ref{eq:dot_2} at corner, edge and internal points, along with using the result from Lemma~\ref{lem:1}, gives us the results in Lemma~\ref{lem:2}. A detailed proof of this has been deferred to Section F in the supplementary material.

\begin{figure}[t]
    \centering
    \includegraphics[width=0.6\columnwidth]{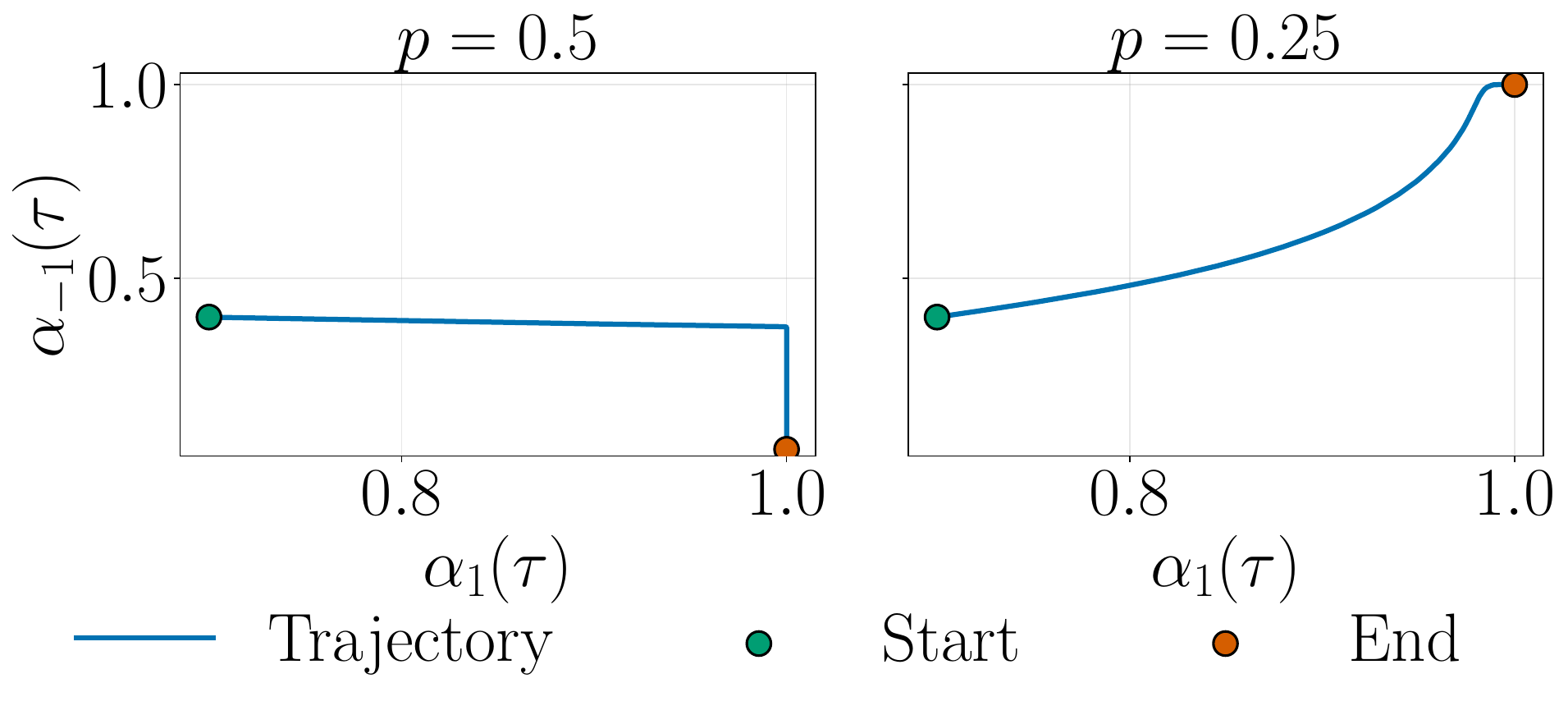}
    \caption{Phase plot of the baseline system after $100,000$ time steps with $p > p^\ast$ (left) and $p < p^\ast$ (right).}
    \label{fig:sim_10}
\end{figure}

\section{Convergence analysis} \label{sec:convergence}
Having identified the equilibrium points, we now proceed to define the necessary and sufficient conditions for the system to converge to them. This analysis reveals how initial parameters govern the system's asymptotic behaviour.

\subsection{Non-Attainability of Simultaneous User Drop-off}
When the RS fails to optimally serve both user classes concurrently, both $\apop$ and $\aniche$ decrease, causing asymptotic convergence to $(0,0)$ (simultaneous user drop-off). Characterising this phenomenon indicates which initial conditions are non-ideal for both user classes.

\begin{theorem}
    \label{thm:4_1}
    Under Assumptions~\ref{ass:1}--\ref{ass:4} and for $p \in (0,1)$, the RS starting from a point $\alpha \in [0,1]^{2} \setminus (0,0)$ cannot asymptotically converge to $(0,0)$.
\end{theorem}

\paragraph{Intuition} For convergence to $(0,0)$, the RS needs both TPR and TNR are both below $0.5$. Thus, its expected performance must be worse than a random guesser for both classes. However, since we update the walk-up rates using $\tstar$ that minimises the expected regularised MSE loss, these misalignments are corrected for, making $(0,0)$ convergence impossible.

\paragraph{Proof Sketch}
\Cref{thm:4_1} is proven by simultaneously solving $\dotpop < 0$ and $\dotniche < 0$. Lemma~\ref{lem:1} helps re-write these inequalities as $(\tstar)^\top \mupop < 0$ and $(\tstar)^\top \muniche > 0$ respectively, which violate the Cauchy-Schwarz inequality. This makes the convergence to $(0,0)$ mathematically impossible; a detailed proof has been provided in supplementary Section G.1.

\begin{figure}[t]
    \centering
    \includegraphics[width=0.6\linewidth]{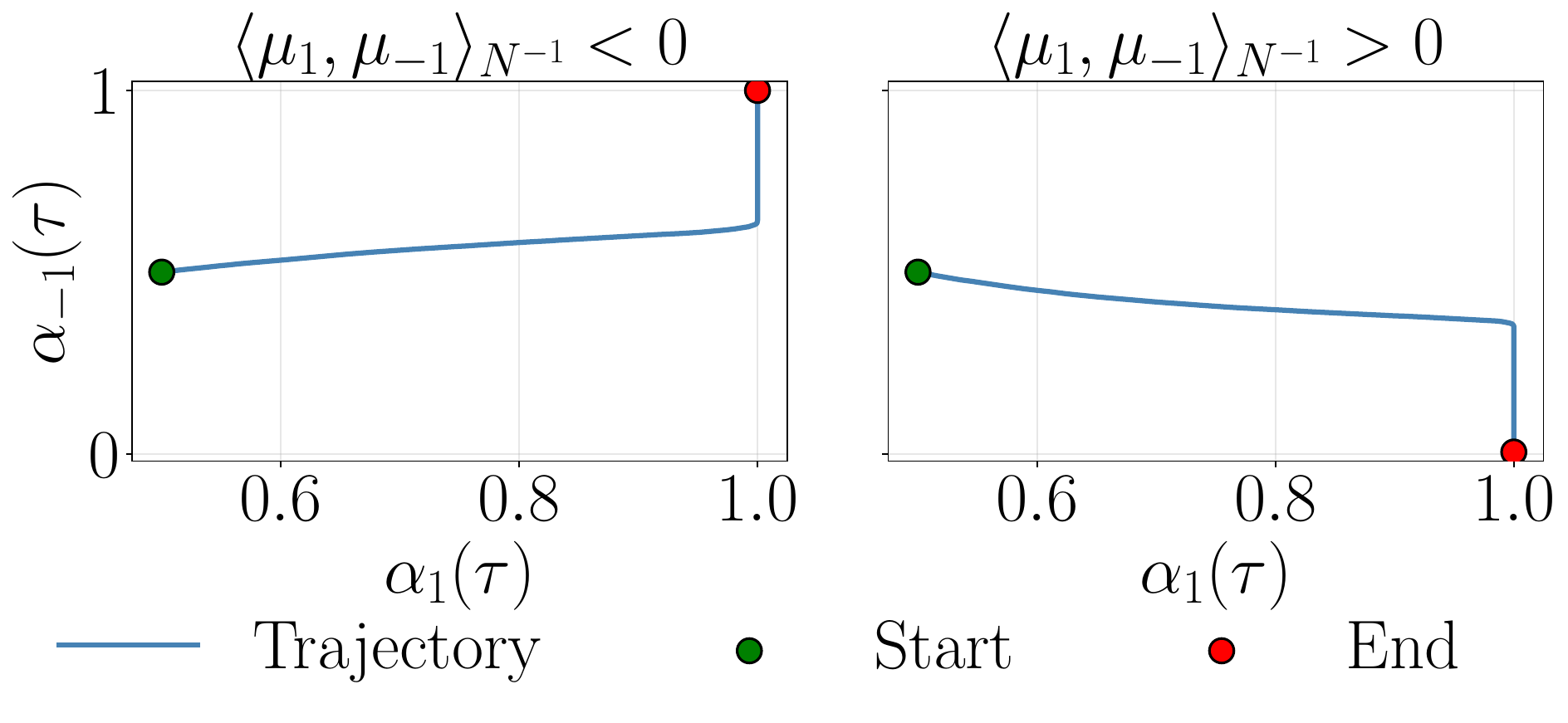}
    \caption{System simulation with engineering setup (left) and the baseline setup (right).}
    \label{fig:sim_43}
\end{figure}

\subsection{Emergence of Popularity Bias}
Popularity bias emerges when the RS favours one user class over another due to its higher representation. In the framework, this leads to the system converging to either $(1,0)$ or $(0,1)$, depending on the favoured class. This section characterises the conditions for the system to converge to $(1,0)$, and the analysis for convergence to $(0,1)$ follows symmetrically (refer to supplementary Section G.4).

\begin{theorem}
    \label{thm:4_2}
    Consider the RS model under Assumptions~\ref{ass:1}--\ref{ass:4} and the mean vectors--$\mupop, \muniche$ such that $\langle \muniche, \mupop \rangle_{A_\alpha^{-1}} > 0$ where $\langle\cdot\rangle_{A_\alpha^{-1}}$ is the inner product admitted by the space transformed by operator $A_\alpha^{-1}$. Then, the system trajectory \textbf{tends} towards $(1, 0)$ if --   
    \begin{equation}
        \label{eq:3}
        p > p^\ast = \frac{\aniche \left\| \muniche \right\|_{A_\alpha^{-1}}^{2} }{\apop\langle \mupop, \muniche \rangle_{A_\alpha^{-1}} + \aniche \left\| \muniche \right\|_{A_\alpha^{-1}}^{2}}
    \end{equation}
    where $\|\cdot\|_{A_\alpha^{-1}}$ is the norm induced in the space transformed by $A_\alpha^{-1}$. Additionally, the system shall \textbf{asymptotically} converge to $(1, 0)$ if and only if 
    \begin{equation}
        \label{eq:3b}
        \frac{\langle \muniche, \mupop \rangle_{B_\alpha}}{\langle \muniche, \mupop \rangle_{A_\alpha^{-1}}} - \frac{\left\|\muniche \right\|_{B_\alpha}^{2}}{\left\| \muniche \right\|_{A_\alpha^{-1}}^{2}} \in \left[ \frac{1}{(p - 1) \aniche}, \frac{1}{p \apop} \right]
    \end{equation}
    where $B_\alpha = A_\alpha^{-1} \left( \spop + \mupop \mupop^\top \right) A_\alpha^{-1}$.
\end{theorem}

\paragraph{Intuition}
For the RS, $p^\ast$ from \Cref{eq:3} acts as a critical threshold on $p$; exceeding which triggers popularity bias, driving the system towards $(1,0)$. This implies the RS can tolerate a deterministic level of class imbalance before favouring the dominant class. Notably, $p^\ast$ depends on both $\alpha$ and $\tstar$ and consequently evolves with time. This is noted in the second statement in \Cref{thm:4_2}. Bounding the variance differences as in \Cref{eq:3b} ensures $p^\ast$ remains non-increasing as the system approaches $(1,0)$, creating an inescapable feedback loop that allows asymptotic convergence.

\paragraph{Proof sketch}
We derive $p^\ast$ in \Cref{eq:3} by simultaneously solving $\dotpop > 0$ and $\dotniche < 0$ using Lemma~\ref{lem:1}. By restricting to the valid cases that obey the Cauchy-Schwarz inequality, we get the expression from \Cref{eq:3}, detailed in supplementary Section G.2. To ensure asymptotic convergence, $p^\ast$ must be non-increasing with respect to $\alpha$ as it approaches $(1,0)$. This requires $\frac{\partial p^\ast}{\partial \apop} \leq 0$ and $\frac{\partial p^\ast}{\partial \aniche} \geq 0$, which when solved together yield \Cref{eq:3b}. This part of the proof has been elaborated in supplementary Section G.3.

\begin{figure*}[t]
    \centering
    \includegraphics[width=\linewidth]{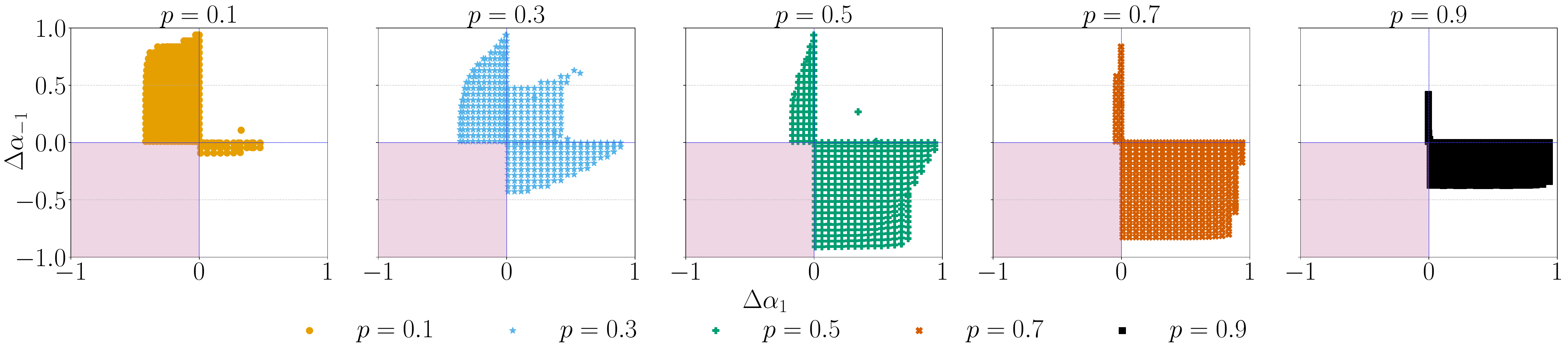}
    \caption{Scatter plot for $p \in \{0.1, 0.3, 0.5, 0.7, 0.9\}$ and $\alpha_i \in [0.01, 0.99]$. Each point corresponds to the mean of 100 MC runs, each for $10,000$ time steps.}
    \label{fig:5_1_scatter}
\end{figure*}

\subsection{Conditions for Symmetric Retention}
Having formalised the non-ideal conditions, we now analyse the ideal case of symmetric user retention, i.e. convergence towards $(1,1)$.

\begin{theorem}
    \label{thm:4_3}
    Consider a system under Assumptions~\ref{ass:1}--\ref{ass:4} such that $ \langle \mupop, \muniche \rangle_{N_\alpha^{-1}} < 0$ where $N_\alpha = p \apop \spop + (1 - p) \aniche \sniche + \lambda I$, then the system shall asymptotically converge to $(1, 1)$ as long as $\sup_{\alpha} \,\, \langle \mupop, \muniche \rangle_{N_\alpha^{-1}} < 0$.
\end{theorem}

\paragraph{Intuition}
The core idea behind \Cref{thm:4_3} is that ``well-separated'' means (a negative inner product) enable the RS to accurately identify user distributions and optimally classify both classes. Enforcing $\sup_{\alpha} \,\, \langle \mupop, \muniche \rangle_{N_\alpha^{-1}} < 0$ ensures this separation of the means across all $\alpha \in [0,1]^{2}$. As a result, this condition holds as the system evolves towards $(1,1)$, thus ensuring asymptotic convergence. Additionally, the inner product is now taken in the space transformed by $N_\alpha^{-1}$ instead of $A_\alpha^{-1}$. The proof sketch highlights that these two inner products are equivalent in our case.

\paragraph{Proof sketch}
To prove \Cref{thm:4_3}, we substitute $\langle \mupop, \muniche \rangle_{N_{\alpha}^{-1}} < 0$ for all $\alpha$ into the expressions for $\dotpop$ and $\dotniche$. Applying Lemma~\ref{lem:1} then yields $\dotpop > 0$ and $\dotniche > 0$. Using simple substitution for $A_\alpha$ can also show that $\langle \mupop, \muniche \rangle_{A_{\alpha}^{-1}}$ and $\langle \mupop, \muniche \rangle_{N_{\alpha}^{-1}}$ share the same sign, rendering them equivalent for our case. The detailed proofs of these claims are in supplementary Section G.5. 

\paragraph{Special case}
A special case of \Cref{thm:4_3} arises when user features follow isotropic Gaussians, where $\spop = \sigma_1^2 I$ and $\sniche = \sigma_{-1}^2 I$. Here, $N$ is a scalar multiple of the identity, meaning $\langle \mupop, \muniche \rangle_{N^{-1}} = \frac{1}{k} \langle \mupop, \muniche \rangle$ for some $k \in \mathbb{R}$. Thus, ensuring $\langle \mupop, \muniche \rangle < 0$ in the original space is sufficient to guarantee asymptotic convergence to $(1,1)$. \\

Although \Cref{thm:4_3} provides sufficient conditions for symmetric user retention, these conditions are not necessary. Convergence to $(1, 1)$ still occurs even when the mean inner product is positive, subject to stricter constraints formalised below.

\begin{theorem}
    \label{thm:4_4}
        Assume the RS model under Assumptions 1-4, initialised at an interior point $\alpha^0=(\alpha_1^0,\alpha_{-1}^0)$. Suppose that $\langle\mu_{-1},\mu_1\rangle_{A_\alpha^{-1}}>0$ for all $\alpha\in[0,1]^2$ such that $\langle\mu_{-1},\mu_1\rangle_{A_\alpha^{-1}}<\langle\mu_{-1},\mu_{-1}\rangle_{A_\alpha^{-1}}$. Also consider functions $f_i(\alpha,p) := (\theta_\alpha^\ast)^\top\mu_i$ and $u_i := A_\alpha^{-1}\mu_i$ for $i\in\{1,-1\}$. Additionally, if the following conditions hold --
        \begin{align*}
            & u_i^\top(\alpha_i S_i+\lambda I)u_{-i}> -\alpha_i(u_i^\top S_{-i}u_i+\frac{\lambda}{\alpha_{-i}}\Vert u_i\Vert ^2) \\
            & \Vert u_1\Vert _{A_\alpha}^2 \ge \langle b_\alpha,S_1 u_1\rangle_{A_\alpha^{-1}} \\
            & \Vert u_{-1}\Vert _{A_\alpha}^2 \ge -\langle b_\alpha,S_{-1} u_{-1}\rangle_{A_\alpha^{-1}} \\
            & \langle u_1,u_{-1}\rangle_{A_\alpha} \le \min(-\langle b_\alpha,S_{-1}u_1\rangle_{A_\alpha^{-1}}, \langle b_\alpha,S_1 u_{-1}\rangle_{A_\alpha^{-1}})
        \end{align*}
    
    where $S_i = \Sigma_i + \mu_i \mu_i^\top$ for $i \in \{ -1, 1\}$. Then there exists $p_1,p_{-1}\in(0,1)$ such that $f_1(\alpha,p_1)=0$ and $f_{-1}(\alpha,p_{-1})=0$, and there exists $p\in(p_1,p_{-1})$ for which the system converges asymptotically to (1, 1).
\end{theorem}

\paragraph{Intuition}
\Cref{thm:4_4} addresses the case where a positive inner product of the means makes it difficult for the RS to distinguish the two classes. We define functions $f_1 (\alpha, p)$ and $f_{-1} (\alpha, p)$, which Lemma~\ref{lem:1} identifies as the update expressions for $\apop$ and $\aniche$, respectively. The list of conditions mentioned ensures two key results -- both functions monotonically increase with $p$, and both transition from negative to positive as $p$ goes from $0 \to 1$. Thus, applying \emph{Intermediate Value Theorem (IVT)} gives $p_1, p_{-1}$ such that $f_1 (\alpha, p_1) = f_{-1} (\alpha , p_{-1}) = 0$. 

From Lemma~\ref{lem:1}, ensuring both $\dotpop$ and $\dotniche$ are positive requires $f_1(\alpha, p) > 0$ and $f_{-1}(\alpha, p) < 0$, which holds for $p \in (p_1, p_{-1})$. Thus, selecting a suitable $p$ ensures asymptotic convergence to $(1,1)$, despite the positively aligned means. 

\paragraph{Proof sketch}
The formal proof involves multiple steps. First, we establish the monotonicity of $f_1, f_{-1}$ by proving their gradients with respect to $p$ are positive. Second, we prove the existence of $p_1, p_{-1}$ by showing $f_i(\alpha, 0) < 0$ and $f_i (\alpha, 1) > 0$ for both $i \in \{-1, 1\}$. Finally, differentiating $f_1, f_{-1}$ with respect to $\alpha$ confirms these trends hold across all states, guaranteeing asymptotic convergence to $(1,1)$. These steps yield the conditions in \Cref{thm:4_4} and the detailed proof is deferred to supplementary Section G.6.

\subsection{Reversal of Trend}
\begin{figure}[t]
    \centering
    \includegraphics[width=0.6\linewidth]{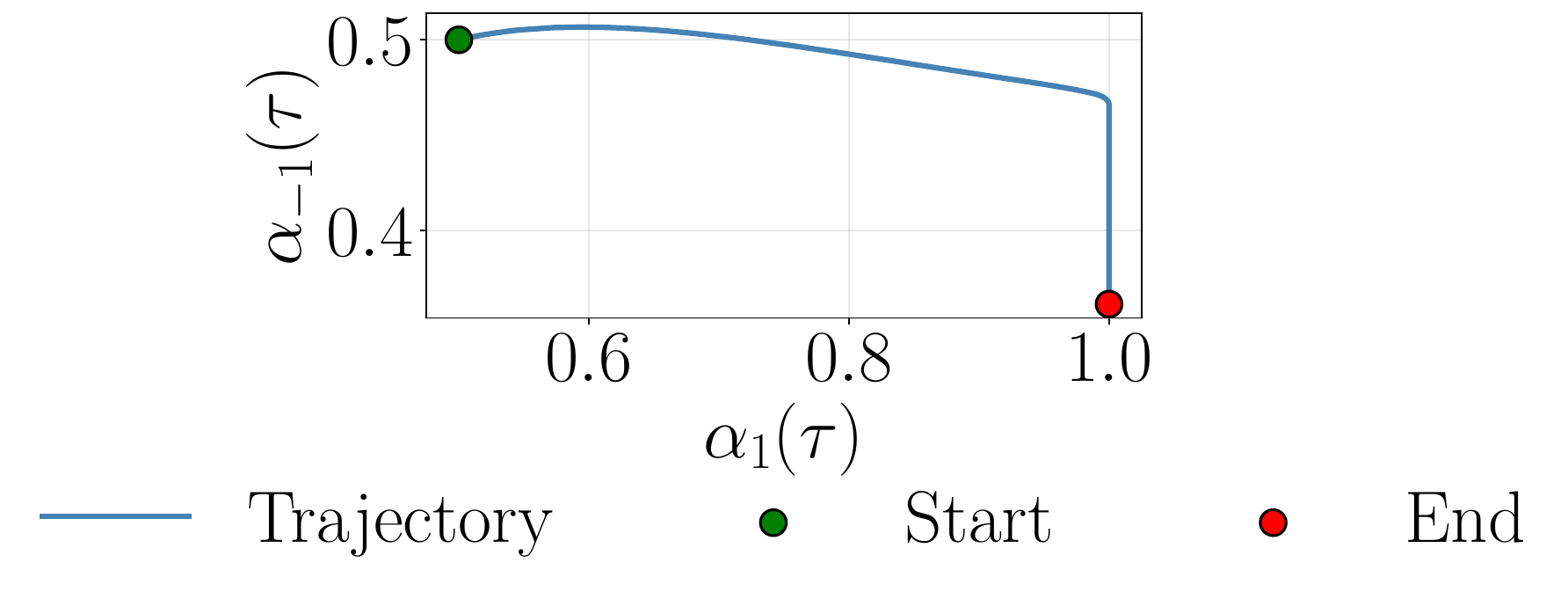}
    \caption{Phase plot of a setup that was simulated for $10,000$ time steps. The plot highlights a reversal of trend from $(1,1)$ to $(1,0)$.}
    \label{fig:sim_11_rev}
\end{figure}

Since $\tstar$ and $p^\ast$ evolve with $(\apop, \aniche)$, it's possible for the system's initial trajectory to eventually reverse. Consider a case with $p = (\apop)_0 = (\aniche)_0 = 0.5$, and the user features distributions $\mupop = \begin{bmatrix} 2 & 5 \end{bmatrix} \,,\, \muniche = \begin{bmatrix} -0.5 & 3 \end{bmatrix} \,,\, \spop = 5I \,,\, \sniche = 2I$. Assuming $\lambda = 1$, the phase plot (\Cref{fig:sim_11_rev}) shows the system initially heading towards $(1, 1)$ before reversing to converge at $(1, 0)$. This is because initially $p^\ast = 0.532 > p$ which pushes the system towards $(1,1)$. However at $((\apop)_t, (\aniche)_t) \approx (0.601, 0.506)$, $p^\ast$ drops to $0.49 < p$, causing the system to converge to $(1, 0)$ as per \Cref{thm:4_2}.

\section{Experiments} \label{sec:expts}
This section focuses on empirically validating the theoretical claims and trends established in \Cref{sec:convergence}, for which we adopt the following baseline simulation setup --
\begin{align*}
    p = 0.5 \,;\, (\apop)_0 = 0.5 \,&;\, (\aniche)_0 = 0.5 \,;\, \lambda = 1 \\
    \mupop = [2, 5] \,;\, \muniche = [-0.5, 2] \,&;\, \spop = 5I \,;\, \sniche = 3I
\end{align*}

The unbiased initialisation $p=(\apop)_0 = (\aniche)_0 = 0.5$ along with $\lambda=1$ guarantees a positive definite $A$. We utilise two-dimensional distributions for visual clarity, selecting means that satisfy \Cref{ass:4} and covariances that inject sufficient noise for both classes. These hyperparameters are varied across experiments to evaluate the different theoretical claims. To account for stochastic user arrivals, we report mean trajectories over \emph{100 Monte Carlo} runs.

Simulations were implemented in Python3 using a \emph{64-core Intel Xeon Silver 4216 CPU} and \emph{four 48GB NVIDIA RTX A6000 GPUs}, complete environment details of which are provided in the README of the code appendix. Finally, because our framework characterises system convergence, we primarily measure the correctness of our claims using system trajectories and final states, explicitly defining alternative metrics for other plots as required.

\begin{figure*}[t]
    \centering
    \begin{subfigure}[b]{0.48\textwidth}
        \centering
        \includegraphics[width=\textwidth]{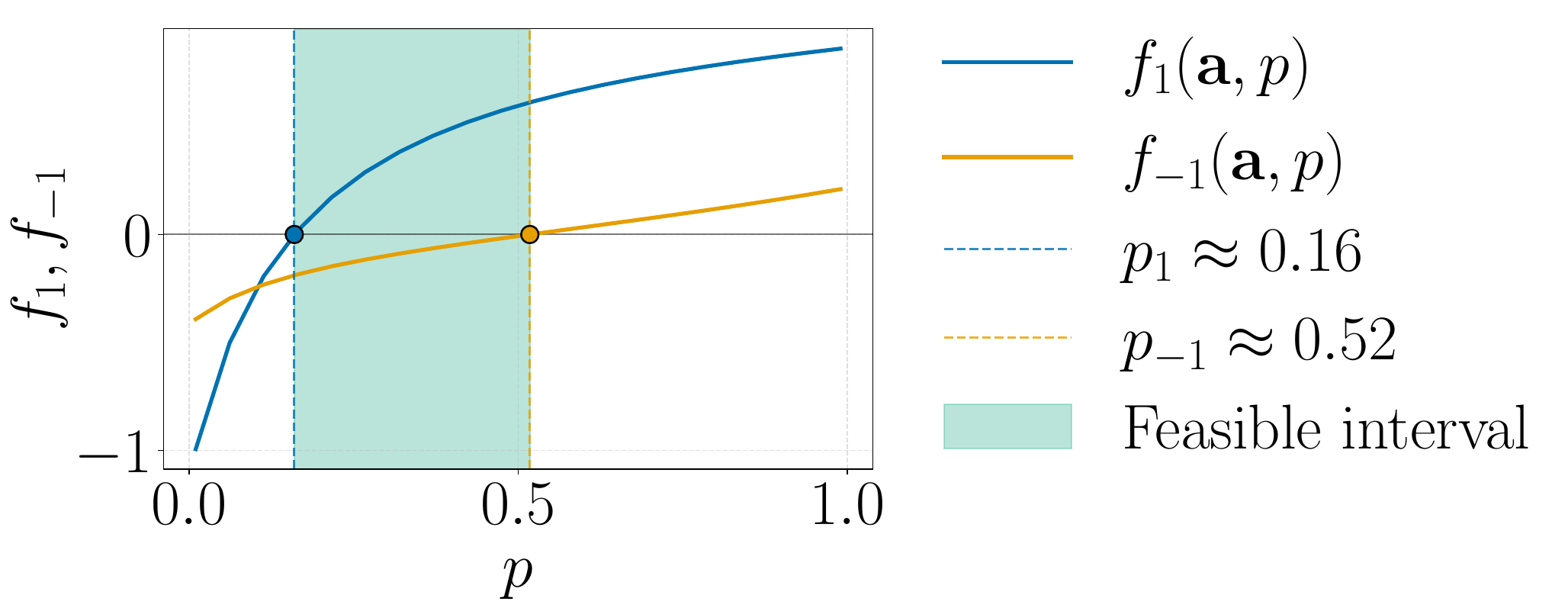}
        \caption{Monotonic nature of $f_1, f_{-1}$.}
        \label{fig:sim_11_44_1}
    \end{subfigure}
    \hfill
    \begin{subfigure}[b]{0.48\textwidth}
        \includegraphics[width=\textwidth]{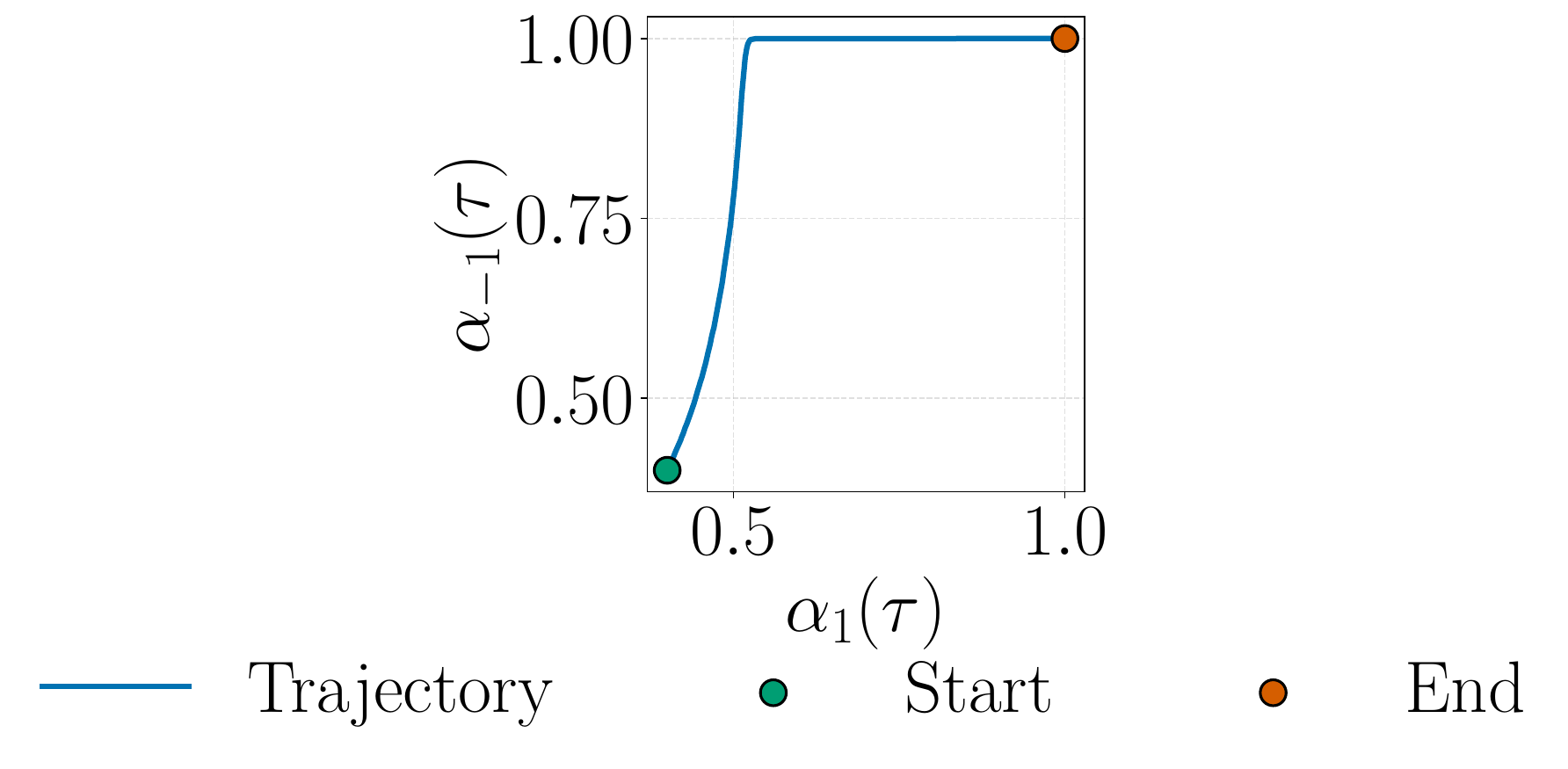}
        \caption{System trajectory converging at $(1,1)$.} 
        \label{fig:sim_11_44_2}
    \end{subfigure}
    \caption{The monotonic nature of $f_1, f_{-1}$ along with their transition from negative to positive as $p$ goes from $0 \to 1$ (left). Presence of $p_1, p_{-1}$ is confirmed and the system trajectory after choosing $p = 0.3 \in (p_1, p_{-1})$ causes the system to converge at $(1,1)$ (right).}
    \label{fig:exp_11_44}
\end{figure*}

\subsection{Verifying \Cref{thm:4_1}}
To empirically verify \Cref{thm:4_1}, scatter plot (\Cref{fig:5_1_scatter}) of the aggregate shifts in walk-up rates $(\Delta\alpha_1, \Delta\alpha_{-1})$ across varying class imbalances $p$ and $\alpha$ was created. The third quadrant (shaded region) of the plot remains empty, confirming the impossibility of simultaneous decreases in $\apop$ and $\aniche$. Additional experiments confirm this finding using phase portraits which deviate away from $(0,0)$ and scatter plots where terminal states never end up at $(0,0)$. Perturbation analysis from the origin indicated an increasing $L_2$ distance between the current state and the origin with time, verifying the unstable nature of $(0,0)$. These observations remain invariant to the values of $\lambda$ and $d$, as confirmed by ablation studies. The experimental details and plots are present in supplementary Section J.1.

\subsection{Verifying \Cref{thm:4_2}}
To validate \Cref{thm:4_2}, the baseline configuration (for which $p^\ast = 0.35$) was simulated using $p = 0.5$ and $p = 0.25$. The phase portraits in \Cref{fig:sim_10} verify that $p = 0.5$ converges to $(1,0)$ unlike $p = 0.25$ which converges towards $(1,1)$, thus verifying \Cref{thm:4_2}. To verify the monotonic nature of $p^\ast$, a heatmap of $p^\ast$ values across different $\alpha$ was created, where the value was monotonically non-increasing as the system approached $(1,0)$, as expected. Finally, ablation studies reveal that higher $\lambda$ or covariance values reduce the value of $p^\ast$ since user parameters get over-shadowed in the system dynamics. Detailed explanations and visualisations have been provided in the supplementary material under Section J.2.

\subsection{Verifying \Cref{thm:4_3}} 
To verify \Cref{thm:4_3}, we engineered a specific setup with $\muniche = [-1, 0]$, $\spop = [[2.0, 0.3], [ 0.3,  1.0]]$ and $\sniche = [[3.0 , -0.2], [-0.2 , 2.0]]$ to ensure negative inner product of the means. Under this setup, the system trajectory converges to $(1,1)$, providing a stark contrast to the simulation with the baseline configuration as shown in \Cref{fig:sim_43}. We also verify the case for isotropic Gaussians using the baseline parameters with $\mupop = [1, 0]$ and $\muniche = [-1, 0]$ such that $\langle\mu_1, \mu_{-1}\rangle < 0$. Simulating the system verifies the sufficiency of this condition for $(1,1)$ convergence. Subsequent ablations confirm that the behaviour remains consistent across different $p$, $d$ and $\lambda$ values. The detailed methodology and plots for these experiments are provided in supplementary Section J.3.

\subsection{Verifying \Cref{thm:4_4}}
The key aspect of \Cref{thm:4_4} is the monotonic nature of $f_i$ for $i \in \{1, -1\}$. To verify that, we engineer a setup with $\mupop = [2, 5]$, $\muniche = [1.5, 0.8]$, $\spop = [[1, 0.3], [0.3, 2]]$ and $\sniche = [[2, -0.2], [-0.2, 1]]$. \Cref{fig:exp_11_44} indicates the monotonic increase of $f_1, f_{-1}$ with respect to $p$ for this setup. It also indicates the change of $f_i$ from negative to positive, implying the existence of $p_i$ for $i \in \{1, -1\}$. Additional experiments consist of checking for negative cases, wherein for both $p < p_1$ and $p > p_{-1}$, the system no longer converges to $(1,1)$. We also plot the gap $\Delta p = p_{-1} - p_1$ and indicate its increases as the system moves towards $(1,1)$, verifying asymptotic convergence. Ablation studies indicate the decrease in $\Delta p$ as $\lambda$ increases, owing to the domination of system dynamics by $\lambda$. The detailed methodology and plots have been deferred to supplementary Section J.4.

\subsection{Experimenting with Real-World Data} \label{sec:rw}
We evaluate the validity of the proposed framework using proprietary interaction logs from a commercial music recommendation platform. The interaction dataset comprises of $\approx {410M}$ user–item interactions along with timestamped clicks, ranked exposure positions, and popularity aggregates across time windows. The simulations performed on this data establish \& explain the presence of popularity bias using the proposed framework. During pre-processing, items with a global popularity in the top \emph{$70$-th percentile} were classified as popular. Consequently, users were classified into three categories based on the fraction of interaction they have with globally popular items---popular-heavy, niche-heavy \& mixed. Popular-heavy and niche-heavy users have $\geq70\%$ of their interaction with popular and niche items respectively.

If a user doesn't log an interaction for \emph{30 consecutive days}, then the user is said to have \emph{churned out}. Else, they are believed to be \emph{retained}. We then analyse the interaction data along two dimensions--\emph{time} and \emph{position}. The time-based analysis checks user retention using metrics like average number of sessions, average number of searches performed etc. The analysis reveals that close to half of the niche-heavy users churn out while only a quarter of the popular-heavy users get churned out. While performing the position/rank analysis, we observe that churned users in general tend to click on items lower on the recommendation list, highlighting the fact that their preferences on average belong to the category of niche items (supplementary Section J.5).

\begin{figure}[t]
    \centering
    \includegraphics[width=0.6\linewidth]{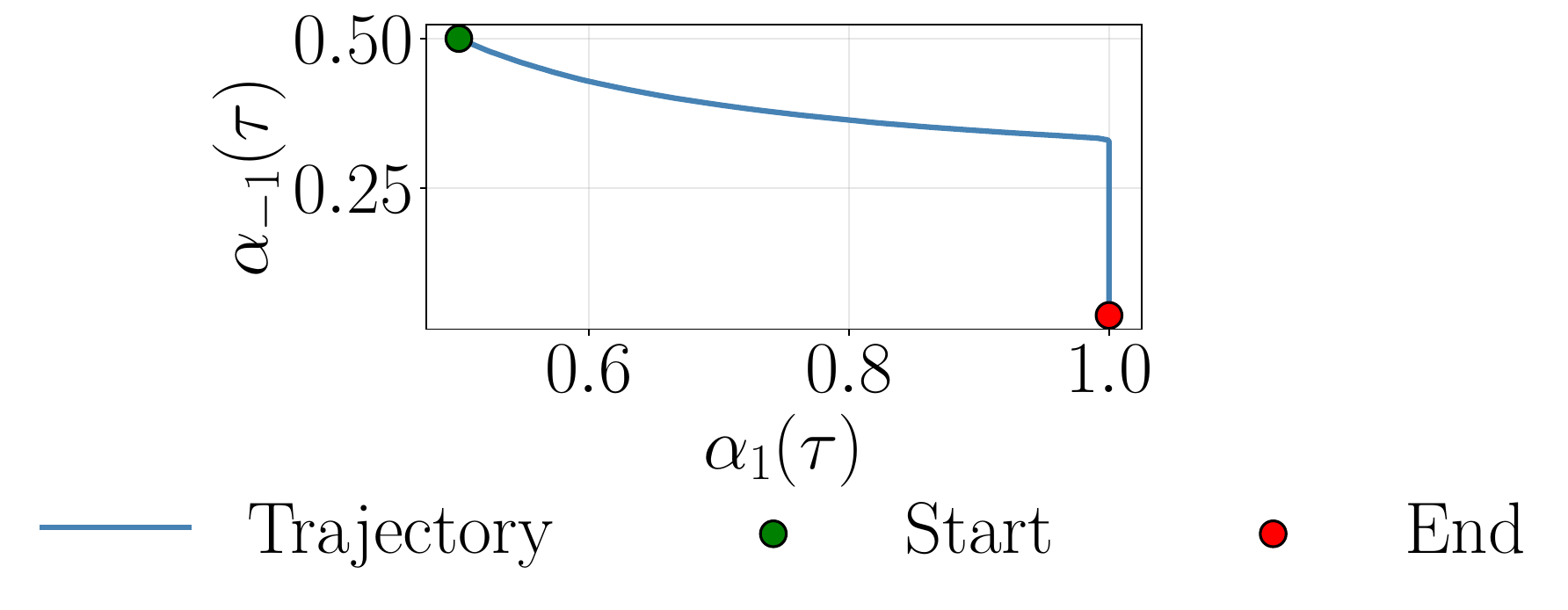}
    \caption{System trajectory for the real-world production logs where the item popularity score is calculated over 14 days.}
    \label{fig:sim_rw}
\end{figure}

With the motivation established, we proceed to fit the interaction data into our dynamic system framework. First, the interaction data is fed to a \emph{Matrix Factorization} model that learns user and item embeddings. Following this, users are classified into \emph{popular} or \emph{niche} based on their preference for popular items. Popular users tend to prefer a globally popular item more than 50\% of the time. Finally, two separate Gaussian distributions are learned over each user group. This gives us the extent of class imbalance, means and covariance matrices for the distribution of the user features. The system satisfies the conditions required by \Cref{thm:4_2} and we expect popularity bias to manifest, which is confirmed after simulating the system for $100,000$ time steps (\Cref{fig:sim_rw}). This provides concrete evidence that the proposed dynamical system does indeed capture bias present in real-world RS.

These results suggest a natural mitigation strategy based on equalizing error rates across the two classes which is similar in spirit to \cite{yu2024fairbalance}. We provide experiments in supplementary Section J.6 showing that this strategy is effective in reducing popularity bias. 

\section{Conclusion \& Future Work} \label{sec:conclusion}
We presented a dynamical system perspective on popularity bias in recommendation systems and, with the help of continuous-time approximations, provided justification for system behaviour. We provide a lower bound on the extent of class imbalance between the users beyond which popularity bias emerges, causing the niche users to drop off asymptotically. We also provide conditions on system parameters for both user classes to be retained in the system. These findings are then verified by simulations -- both on synthetic and real-world data (obtained from a commercial music-recommendation platform). These observations allow for a natural equalised odds mitigation, which has empirically been shown to be effective for the proposed framework.

Future work may extend this framework to generalise over any convex loss function and not just the MSE loss (supplementary Section C). Additionally, relaxation of assumptions on the user feature distributions can lead to more generalised results. More broadly, we believe that the proposed framework can be extended to understand and justify multi-user class dynamics as well (supplementary Section B).

\bibliographystyle{unsrt}  
\bibliography{references}  






\end{document}